\documentclass[aps,pra,twocolumn,amssymb,amsmath,%
    twoside,%a4paper,floatfix,%
    superscriptaddress,noshowpacs,noshowkeys]{revtex4}

\usepackage{graphicx}

\newcommand*{\ket}[1]{|{#1}\rangle}
\newcommand*{\bra}[1]{\langle{#1}|}

\newcommand*{\eref}[1]{Eq.~(\ref{#1})}
\newcommand*{\Eref}[1]{Eq.~(\ref{#1})}
\newcommand*{\sref}[1]{Sec.~\ref{#1}}
\newcommand*{\Sref}[1]{Sec.~\ref{#1}}

\newcommand*{\sxa}{\sigma_{X,A}}
\newcommand*{\spa}{\sigma_{P,A}}
\newcommand*{\sxl}{\sigma_{X,L}}
\newcommand*{\spl}{\sigma_{P,L}}
\newcommand*{\sta}{\sigma_\eta}

\DeclareMathOperator{\Span}{Span}

\newcommand*{\textint}{{\text{int}}}

\begin{document}
\title{Continuous variable versus EIT-based quantum memories}
\author{Z. Kurucz}
\affiliation{Department of Physics, University of Kaiserslautern,
  D-67663 Kaiserslautern, Germany}
\affiliation{Department of Nonlinear and Quantum Optics, Research
  Institute for Solid State Physics and Optics, Hungarian Academy of
  Sciences, PO Box 49 H-1525 Budapest, Hungary}
\author{M. Fleischhauer}
\affiliation{Department of Physics, University of Kaiserslautern,
  D-67663 Kaiserslautern, Germany}

\date{Dec.\ 14, 2007}

\begin{abstract}
  We discuss a general model of a quantum memory for a single light
  mode in a collective mode of atomic oscillators.  The model includes
  interaction Hamiltonians that are of second order in the canonical
  position and momentum operators of the light- and atomic oscillator
  modes.  We also consider the possibility of measurement and
  feedback.  We identify an interaction Hamiltonian that leads to an
  ideal mapping by pure unitary evolution and compare several schemes
  which realize this mapping using a common continuous-variable
  description.  In particular we discuss schemes based on the
  off-resonant Faraday effect supplemented by measurement and feedback
  and proper preparation of the atoms in a squeezed state and schemes
  based on off-resonant Raman coupling as well as electromagnetically
  induced transparency (EIT).
\end{abstract}
%\pacs{}%
%\keywords{}%
\maketitle

%%%%%%%%%%%%%%%%%%%%%%%%%%%%%%%%%%%%%%%%%%%%%%%%%%%%%%%%%%%%%%%%%%
%                                                                %
%                           Section                              %
%                                                                %
%%%%%%%%%%%%%%%%%%%%%%%%%%%%%%%%%%%%%%%%%%%%%%%%%%%%%%%%%%%%%%%%%%
\section{Introduction}
\label{sec:intro}

One of the key elements of quantum communication schemes
\citep{prl81p5932, nat414p413} and network quantum computers
\citep{prl78p3221, quantph9704012} is a high-fidelity, reversible
quantum memory.  In recent years substantial progress has been made in
the design \cite{%
  prl85p5643, qicv231, prl95e053602, pra73e022331, %
  pra73e062329, pra74e011802, % (Faraday theory)
  pra62e033809, pra69e043810, pra71e043801, % (Raman theory)
  prl84p4232, prl84p5094, pra65e022314, pra72e022327, pra76e033804% (EIT theory)
} and experimental realization \cite{%
  prl89e057903, nat432p482, sci304p270, % (Faraday exp)
  sci301p196, % (Raman exp)
  sci306p663, nat409p490, prl86p783, %
  prl89e067901, nat426p638, prl92e203602, nat438p837% (EIT exp)
} of such memories based on ensemble of atoms or other quantum
radiators as storage unit and photons as information carrier. In
particular two distinct approaches to a quantum light-matter interface
have been put forward: one making use of the off-resonant Faraday
effect to transfer a qubit encoded in the polarization state of a
light field to a macroscopic atomic spin of the atomic ensemble
\cite{prl85p5643, qicv231, prl95e053602, pra73e022331, %
  pra73e062329, pra74e011802, % (Faraday theory)
  prl89e057903, nat432p482, sci304p270% (Faraday exp)
}, and another one using Raman interactions and electromagnetically
induced transparency (EIT) \cite{%
  pra62e033809, pra69e043810, pra71e043801, % (Raman theory)
  prl84p4232, prl84p5094, pra65e022314, pra72e022327, %
  pra76e033804, % (EIT theory)
  sci301p196, % (Raman exp)
  sci306p663, nat409p490, prl86p783, prl89e067901, %
  nat426p638, prl92e203602, nat438p837% (EIT exp)
}. Currently substantial work is devoted to the optimization of these
schemes which is crucial for potential large-scale implementations.
An obstacle in this effort is the rather different theoretical
frameworks used to describe these approaches. We here put forward a
common description of the Faraday and Raman schemes, which can be used
to compare their advantages and drawbacks.

We begin with a review of general properties of a single-mode quantum
memory in terms of continuous light and matter variables recovering
ideal Hamiltonians for a purely unitary realization of the map.  We
then discuss the possibility to realize an ideal one-way map from
light to atoms using Hamiltonians that are not equivalent to the ideal
one, such as the Faraday interaction Hamiltonian, by means of
measurement and feedback techniques as well as proper state
preparation of the atomic ensemble \cite{qicv231, nat432p482}. It is
shown that the rather demanding conditions on measurement and state
preparation can be substantially reduced in double-pass configurations
\cite{pra74e011802} and can be totally eliminated in a triple-pass
scheme.

We then discuss physical implementations of the different mapping
approaches starting from the $J=-1/2 \leftrightarrow J=+1/2$ scheme of
Ref.~\cite{nat432p482} realizing a Faraday coupling.  We then show
that an off-resonant Raman coupling in a $J=n \leftrightarrow J = m$
configuration allows to implement both the Faraday coupling
Hamiltonian and the ideal mapping Hamiltonian.  Finally we discuss the
quantum memory based on EIT \cite{prl84p5094} and show that this
scheme corresponds to an ideal mapping Hamiltonian when a proper spin
polarized (but not squeezed) initial state of the atomic ensemble is
considered.

The paper is organized as follows.  In \sref{sec:Hamiltonians}, we
consider a simple model for quantum memories in terms of atomic and
light field quadrature variables.  In \sref{sec:memphys}, we discuss
physical systems that can be used to realize Faraday-type,
off-resonant Raman-type, and EIT interaction Hamiltonians.
\Sref{sec:conclusions} summarizes our results.

%%%%%%%%%%%%%%%%%%%%%%%%%%%%%%%%%%%%%%%%%%%%%%%%%%%%%%%%%%%%%%%%%%
%                                                                %
%                           Section                              %
%                                                                %
%%%%%%%%%%%%%%%%%%%%%%%%%%%%%%%%%%%%%%%%%%%%%%%%%%%%%%%%%%%%%%%%%%
\section{Realizations of the ideal map using unitary evolution,
measurement, and feedback}
\label{sec:Hamiltonians}

We consider an abstract model of a reversible
memory for the quantum state of a light mode in an ensemble of atoms.
The light mode (system~$L$) is described in terms of the canonical
quadrature variables $\hat X_L$ and $\hat P_L$. We assume that the
quantum memory (system~$A$) to which we intend to transfer the quantum
state can be described by a similar set of continuous variables $\hat
X_A$ and $\hat P_A$ with $[\hat X_A, \hat P_A]= i$, ($\hbar =1$). The
latter is the case, e.g., for a large ensemble of initially polarized
spins, if the excitation probability of each individual spin is small.
The time evolution of the two quantum systems can be described in the
Heisenberg picture by a map that connects the dynamical variables of the
systems at some initial time to those at a final time $t$:
% \begin{equation}
  $(\hat X_A(0), \hat P_A(0), \hat X_L(0), \hat P_L(0)) \mapsto
  (\hat X_A(t), \hat P_A(t), \hat X_L(t), \hat P_L(t))$.
% \end{equation}
For an ideal quantum memory, we require the mapping to be linear in
the quadrature variables and complete in the sense that the variables
of one subsystem are mapped only to those of the other.  That is,
employing the vector notation
\begin{equation}
  \label{eq:y}
  \hat{\mathbf y} \equiv 
  (\hat X_A, \hat P_A, \hat X_L, \hat P_L)^T,
\end{equation}
the map has the compact form
\begin{equation}
  \hat{\mathbf y}_{\rm out}
  = {\mathbf M}\, \hat{\mathbf y}_{\rm in}
  \quad \mbox{with} \quad \mathbf M =
  \left(\begin{array}{cc}
      0 & \mathbf M_1 \\
      \mathbf M_2 & 0
    \end{array}\right),
\end{equation}
with ${\mathbf M}_i$ being 2 $\times$ 2 symplectic, real matrices.
The matrices need to be symplectic in order to conserve commutation relations.

To implement this map we consider a Hamiltonian evolution that may be
supplemented by measurement and feedback processes. To ensure the
linearity of the map, the Hamiltonian should be of at most second
order in the quadrature variables (or in the corresponding
annihilation and creation operators).  Specifically, we consider pure
harmonic oscillators with quadratic interaction between them:
\begin{eqnarray}
  \label{eq:H}
  \hat H &=& \hat H_A + \hat H_L + \hat H_\textint
  ,\\
  \label{eq:H_A}
  \hat H_A &=& 
  \frac{\omega_A}{2}\left(\hat X_A^2 + \hat P_A^2\right) 
  = \omega_A \left(\hat a_A^\dagger \hat a_A +\frac12\right)
  ,\\
  \label{eq:H_L}
  \hat H_L &=& 
  \frac{\omega_L}{2}\left(\hat X_L^2 +\hat P_L^2\right)
  = \omega_L \left(\hat a_L^\dagger \hat a_L +\frac12\right)
  ,\\
  \label{eq:H_int}
  \hat H_\textint &=&
  p \hat X_A \hat X_L + q \hat X_A \hat P_L +
  r \hat P_A \hat X_L + s \hat P_A \hat P_L
  ,
\end{eqnarray}
where $\omega_A$ and $\omega_L$ are the
oscillator frequencies and the real parameters $p$,~$q$, $r$, and~$s$
characterize the interaction.  These parameters may have explicit time
dependence.

%%%%%%%%%%%%%%%%%%%%%%%%%%%%%%%%%%%%%%%%%%%%%%%%%%%%%%%%%%%%%%%%%%
\subsection{Purely unitary evolution}
\label{sec:pureunitary}
%%%%%%%%%%%%%%%%%%%%%%%%%%%%%%%%%%%%%%%%%%%%%%%%%%%%%%%%%%%%%%%%%%

Let us first consider the question under what conditions an ideal quantum
memory map can be realized by pure unitary evolution
\cite{pra67e042314, pra68e022304}.  For this we
assume that the free Hamiltonian $\hat H_0=\hat H_A +\hat H_L$
commutes with the interaction Hamiltonian,
\begin{equation}
  \label{H0-H1-commute}
  [\hat H_0(t), \hat H_\textint(t)] = 0 .
\end{equation}
With this restriction, the free Hamiltonian can be eliminated from the
equations of motion.  Indeed, let us express operators of the
Heisenberg picture ($\hat A_H(t)$) in the frame rotating according to the free
Hamiltonian: $\hat A(t) \equiv \hat U_0(t) \hat A_H(t) \hat
U_0^\dag(t)$, where $\hat U_0(t)$ denotes the unitary operator of the
interaction-free time evolution.  Thus, the equation of motion for
operators in the rotating frame reads
\begin{equation}
  \label{eq-rotating-frame}
  \frac d{dt} \hat A(t)
  =i [H_\textint (t),\hat A(t)].
\end{equation}
If we further assume that the interaction Hamiltonian commutes at
different times, $[\hat H_\textint (t), \hat H_\textint (t')] =0$,
then the time evolution operator corresponding to the interaction can
be written in an exponential form with no time ordering necessary.
This assumption together with \eqref{H0-H1-commute} necessarily
implies exact resonance between the atomic and light systems
($\omega_A = \omega_L = \omega$) and that the interaction Hamiltonian
is of the form $\hat H_\textint (t) = \alpha(t) \hat H_1$ with
\begin{equation}
  \label{eq:H_1}
  H_1 \equiv \sin\xi (\hat X_A \hat X_L + \hat P_A \hat P_L)
  + \cos\xi (\hat P_A \hat X_L - \hat X_A \hat P_L),
\end{equation}
that is, $\hat H_\textint$ can have explicit time dependence only
through $\alpha(t)$.  Then \eref{eq-rotating-frame} can formally be
solved using the Baker-Campbell-Hausdorff formula
\begin{equation}
  \label{eq:baker-haus}
  \hat A(t)=\sum_{n=0}^\infty \frac1{n!} \left(
    \int_{0}^t d\tau \, \alpha(\tau)\right)^n \, \hat A_n,
\end{equation}
where $\hat A_n=(-i)^n[[[\hat A(0),\hat H_1],\hat H_1],\dots]$ is
proportional to the $n$-fold commutator of $\hat A(0)$ with $\hat
H_1$.  It is easy to see that the commutators of the quadrature
variables $\hat{\mathbf y}$ with the interaction Hamiltonian
(\ref{eq:H_1}) are linear in the same set of quadratures, namely, we
have
\begin{equation}
  \left[\hat{\mathbf y},\hat H_1\right] = 
  i\, {\mathbf C} \, \hat{\mathbf y}
  \quad\mbox{and}\quad 
  \hat {\mathbf y}_n = \mathbf C \, \hat{\mathbf y}_{n-1}
  =\mathbf C^n\, \hat{\mathbf y},
\end{equation}
where
\begin{equation}
  \label{eq:C-matrix}
  {\mathbf C} = \left(
    \begin{array}{cc}
      \mathbf 0 & -{\mathbf R}^{-1}\\
      {\mathbf R} & \mathbf 0
    \end{array}
  \right)
  \quad \mbox{and} \quad
  {\mathbf R} = \left(
    \begin{array}{cc}
      \cos\xi & -\sin\xi \\
      \sin\xi &  \cos\xi
    \end{array}
  \right).
\end{equation}
Using the series \eqref{eq:baker-haus} and recognizing that
$\mathbf C^2 = -\mathbf 1$, we find the linear relation between the
quadratures at time $t$ and the initial time
\begin{equation}
  \hat {\mathbf y}(t) 
  % = \exp\big[\Phi(t) \mathbf C\big] \hat{\mathbf y}(0),\\
  = \left[\cos\Phi(t) \, {\mathbf 1} +
    \sin\Phi(t) \left(\begin{array}{cc}
        {0} & -\mathbf R ^{-1} \\
        \mathbf R & {0}\end{array}\right)
  \right]\hat{\mathbf y}(0),
\end{equation}
with
\begin{equation}
\Phi(t)= \int_{0}^t d\tau\, \alpha(\tau).
\end{equation}
We see that perfect quantum memory mapping is achieved after an
interaction time $T$ such that the area of coupling is $\Phi(T) = (2
n+1)\pi/2$, (with %$n=0$,~$\pm 1$, $\pm 2$,~\dots
$n\in\mathbb Z$).  In this case one has
\begin{equation}
  \hat{\mathbf y}(T)= \pm\left(
    \begin{array}{cc}
      {0} & -{\mathbf R}^{-1} \\
      {\mathbf R} & {0}
    \end{array}\right)\, \hat{\mathbf y}(0).
\end{equation}

Regarding the possibility of physical realizations of the interaction
Hamiltonian \eqref{eq:H_1}, two different values of $\xi$ are of
particular importance.  For $\xi=0$, we have
\begin{equation}
  \label{eq:Hideal1}
  \hat H_\textint (t) = \alpha(t)
  \left(\hat P_A \hat X_L - \hat X_A \hat P_L\right).
\end{equation}
If the envelope $\alpha(t)$ is chosen such that $\Phi(T)=\pi/2$ then
we arrive at an ideal quantum memory map
\begin{gather}
  \begin{aligned}
  \label{eq:ideal1}
    \hat X_A (T) &= \hat X_L, &
    \hat P_A (T) &= \hat P_L, \\
    \hat X_L (T) &= -\hat X_A, & 
    \hat P_L (T) &= -\hat P_A.
  \end{aligned}
\end{gather}
We can also consider the storing process in which the $\hat X$ and
$\hat P$ quadratures of the memory system are interchanged with
respect to the previous transformation.  This corresponds
to $\xi=\pi/2$ and the interaction Hamiltonian
\begin{equation}
  \label{eq:Hideal2}
  \hat H_\textint (t) = \alpha(t)
  \left(\hat X_A \hat X_L + \hat P_A \hat P_L\right)
\end{equation}
leads to the map
\begin{gather}
 \begin{aligned}
  \label{eq:ideal2}
    \hat X_A (T) &= \hat P_L, &
    \hat P_A (T) &= -\hat X_L, \\
    \hat X_L (T) &= \hat P_A, & 
    \hat P_L (T) &= -\hat X_A,
  \end{aligned}
\end{gather}
which is again the map of an ideal quantum memory.

%%%%%%%%%%%%%%%%%%%%%%%%%%%%%%%%%%%%%%%%%%%%%%%%%%%%%%%%%%%%%%%%%%
%                                                                %
%                           SubSection                           %
%                                                                %
%%%%%%%%%%%%%%%%%%%%%%%%%%%%%%%%%%%%%%%%%%%%%%%%%%%%%%%%%%%%%%%%%%
\subsection{Single-pass scheme with feedback and initial spin squeezing}
\label{sec:1pass}

A quantum memory for light, which is not based entirely on unitary
evolution but is rather an approximate simulation of the Hamiltonian
\eqref{eq:Hideal2}, was proposed and experimentally demonstrated
in~\cite{nat432p482}.  The light-matter interaction there is due to
the Faraday effect and is described by the Hamiltonian
\begin{equation}
  \label{eq:1pass}
  \hat H_0=\text{const.},
  \qquad \hat H_1 = \hat P_A \hat P_L.
\end{equation}
See \sref{sec:Faraday} for a possible derivation of \eqref{eq:1pass}.
In this case the matrix ${\mathbf C}$ cannot be represented in the
form \eqref{eq:C-matrix} and the unitary evolution will simply shift
the position operators by an amount proportional to the momentum of
the other system as well as to interaction time $t$, while the momenta
are constants of motion:
\begin{gather}
  \label{eq:1passevol}
  \begin{aligned}
  \hat X_A (t) &= \hat X_A + t \hat P_L, &
  \hat P_A (t) &= \hat P_A,\\
  \hat X_L (t) &= \hat X_L + t \hat P_A, &
  \hat P_L (t) &= \hat P_L.
  \end{aligned}
\end{gather}
One recognizes that only the momentum quadrature of the light mode
is transferred to the atomic ensemble. To also map the
position quadrature to the ensemble the unitary
evolution was complemented in Ref.~\cite{nat432p482}
by a homodyne measurement of the outgoing light
quadrature $\hat X_L$ and the measurement result $x$ was fed back by
applying a momentum displacement of $-x/t$ on system $A$.
As a consequence of the measurement, one can formally write the c-number
$x$ in place of $\hat X_L(t)$ and rearrange (\ref{eq:1passevol}) to
conclude that after measurement and feed-back:
\begin{equation}
  \label{eq:1pass+feedback}
  \hat X_A^{\mathrm{mem}} = \hat X_A + t \hat P_L, 
  \qquad
  \hat P_A^{\mathrm{mem}} = -\frac 1t \hat X_L.
\end{equation}
If the atomic ensemble were initially prepared in a position
eigenstate, i.e., in an infinitely squeezed state, the operator $\hat
X_A$ in (\ref{eq:1pass+feedback}) could be replaced by a c-number. The
resulting map would in this case ideally transfer the complete state
of light to the atomic ensemble.

To verify these statements in a more rigorous way, we calculate the
state of the quantum memory after the storage and the storage fidelity
in terms of Wigner functions.  The atomic and light systems are
initially disentangled, so the two-particle Wigner function is of the
product form
\begin{equation}
  \label{eq:W0}
  W_0(x_A,p_A; x_L,p_L) = W_A(x_A,p_A)\, W^{\mathrm{in}}(x_L,p_L),
\end{equation}
%
%
%where the Wigner functions are defined as in~\cite{physrep106p121}.
After the transformation (\ref{eq:1passevol}), the new state is given by
\begin{equation}
  \label{eq:W1}
  W'(x_A,p_A; x_L,p_L) = W_0(x_A-tp_L,p_A; x_L-tp_A,p_L).
\end{equation}

Then the quadrature $\hat X_L'$ of the outgoing light is measured.
For an ideal measurement, the projection corresponding to the outcome
$x$ is $\hat \openone_A \otimes \hat \Pi_x$, where $\hat \Pi_x = \ket
x_{LL}\bra x$ and its Wigner function is $\Pi_x(x_L',p_L') =
\delta(x_L'-x)$.  The (unnormalized) conditional atomic state can be
obtained by a von Neumann projection.  The feedback is described by a
shift in the atomic momentum $p_A' = p_A'' + x/t$ and thus the 
state of the memory conditioned on the measurement result $x$ 
and after feedback reads:
\begin{multline}
  \label{eq:Wmemx}
  W^{\mathrm{mem}}_x (x_A',p_A'')  = \int dx_L' dp_L'\,  
  W_A(x_A'-tp_L', p_A''+x/t) 
  \\\times  W^{\mathrm{in}}(x_L'-tp_A''-x, p_L') \, \Pi_x(x_L', p_L').
\end{multline}
The norm of (\ref{eq:Wmemx}) gives the probability distribution $P(x)$
of the measurement outcome $x$.  In general, $P(x)$ has explicit
dependence on the unknown input state $W^{\mathrm{in}}(x_L, p_L)$.
This means that the measurement yields information about the input
state of the light which is, therefore, distorted.  However, in the
special case of an infinitely squeezed initial atomic state
$W_A(x_A,p_A) = \delta(x_A-x_0)$, we have a uniform probability
distribution $P(x)=1/t$ and no information about the initial light
state is obtained by the measurement.  In this ideal case, the final
state is
\begin{equation}
  \label{eq:Wmemideal}
  W^{\mathrm{mem}} (x,p) = W^{\mathrm{in}}(-tp, (x-x_0)/t)
\end{equation}
that corresponds to an ideal quantum memory mapping
regardless of the outcome $x$.

In reality, however, it is impossible to prepare the atomic ensemble
in a position eigenstate, i.e., in an infinitely squeezed state. 
Furthermore, the measurement will be imperfect in general. In the 
followings, we discuss the effects of finite initial squeezing and
finite detection efficiency. For this we replace the Wigner function
$\Pi_x$ used above with the corresponding expression for a finite
detection efficiency~\cite{pra48p4598}
\begin{eqnarray}
  \nonumber
  \hat \Pi_{x,\sta} &=& ({2\pi \sta^2})^{-1/2}
  \int dy\, e^{-\frac{(y-x)^2}{2\sta^2}} \ket y_{LL}\bra y,
  \\\label{eq:Pinoisy}
  \Pi_{x,\sta} (x_L',p_L') &=& ({2\pi \sta^2})^{-1/2}
  \exp\left\{-\frac{(x_L'-x)^2}{2\sta^2}\right\}.
\end{eqnarray}
Here $\sta$ characterizes the resolution of the position measurement,
with $\sta=0$ corresponding to a noiseless, perfect measurement, while
a typical experimental value is in the order of $\sta^2 = 2.5 \times
10^{-3}$.  Averaging over the measurement outcome $x$ gives the memory
state
\begin{multline}
  \label{eq:Wmemnoisy}
  W^{\mathrm{mem}} (x,p) = \int dx' dp' dx'' \,
  W_A (x',p') \\ \times
  W^{\mathrm{in}} (-tp + \sta x'', (x-x')/t) \, \Pi_{0,1}(x'',0).
\end{multline}

\begin{figure}
  \centering
  \includegraphics[width=\hsize]{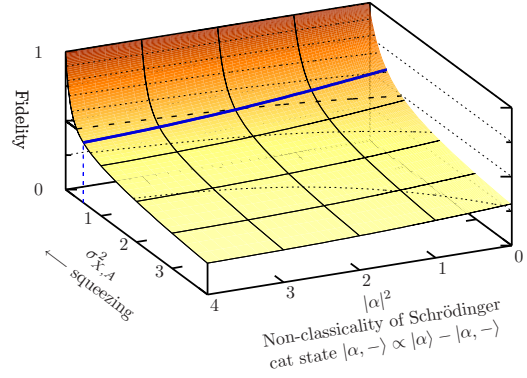}
  \caption{%(color online)
    Fidelity of storing the odd Schr\"odinger cat states
    $\ket{\alpha,-} \propto \ket\alpha - \ket{-\alpha}$ (with real
    $\alpha$) in the one-pass scheme of~\cite{nat432p482}.  The atomic
    ensemble is initially prepared in a spin squeezed state: the
    smaller the position variance $\sxa$, the higher the squeezing.
    Thick line corresponds to the coherent spin state with no
    squeezing and it reaches the classical limit of fidelity 0.5
    (thick dashed contour line) at about $|\alpha|^2=2.0$.  Since the
    measurement alters the input state, the scheme cannot efficiently
    store superpositions of states with very different momenta.}
  \label{fig:cat-fidel}
\end{figure}

Let us consider  as the initial atomic state a Gaussian spin squeezed state
\begin{equation}
  \label{eq:WAreal}
  W_A(x,p) = \frac1{2\pi \sxa \spa}
  e^{-\frac{(x-x_0)^2}{2\sxa^2}
    -\frac{(p-p_0)^2}{2\spa^2} }.
\end{equation}
In the experiment of~\cite{nat432p482}, it was actually a coherent
spin state with $\sxa^2 = \spa^2 = \frac12$.  The imperfect
measurement (\ref{eq:Pinoisy}) then results the state
\begin{multline}
  \label{eq:Wmemreal}
  W^{\mathrm{mem}}(x,p) 
%  =\int dx\, W^{\mathrm{mem}}_x (x_A,p_A)
  =\frac1{2\pi \sta \sxa} 
  \int dx' dp' \, 
  e^{-\frac{x'^2}{2\sta^2} - \frac{t^2 p'^2}{2\sxa^2}}
  \\\times
  W^{\mathrm{in}}(-tp+x',(x-x_0)/t+p')
\end{multline}
for the atomic quantum memory.  Compared to the ideal memory state
(\ref{eq:Wmemideal}), we have higher uncertainties both in the $X$ and
$P$ quadratures due to the noisy measurement and the imperfect initial
state preparation, respectively.  The average fidelity of the storage
process can be calculated from the overlap of the ideal
(\ref{eq:Wmemideal}) and real (\ref{eq:Wmemreal}) output states,
\begin{equation}
  \label{eq:Fx}
  F   = 2\pi \int dx dp\, W^{\mathrm{in}}(-tp,(x-x_0)/t)\,  
  W^{\mathrm{mem}} (x, p).
\end{equation}
For a Gaussian state of the input light field
\begin{equation}
  \label{eq:WLGaus}
  W^{\mathrm{in}}(x,p) = \frac1{2\pi \sxl \spl}
  e^{ -\frac{(x-x^{\mathrm{in}})^2}{2\sxl^2} 
  -\frac{(p-p^{\mathrm{in}})^2}{2\spl^2}},
\end{equation}
the fidelity reads
\begin{equation}
  F = [(\sxa^2/t^2 + 2\spl^2) (\sta^2 + 2\sxl^2)]^{-1/2},
\end{equation}
which is about 82\% for a coherent spin state and coherent light input
($\sxa^2=\sxl^2=\spl^2=1/2$, $\sta=0$, and $t=1$), as in
Ref.~\cite{nat432p482}.  However, direct calculation shows that the
fidelity of storage quickly decreases for highly nonclassical states
like Schr\"odinger cat states (see Fig.~\ref{fig:cat-fidel}).

%%%%%%%%%%%%%%%%%%%%%%%%%%%%%%%%%%%%%%%%%%%%%%%%%%%%%%%%%%%%%%%%%%%%%%%%%%
%%%%%%%%%%%%%%%%%%%%%%%%%%%%%%%%%%%%%%%%%%%%%%%%%%%%%%%%%%%%%%%%%%%%%%%%%%

We conclude that the single-pass feedback technique provides the
perfect quantum memory mapping if and only if the atomic ensemble is
initially prepared in an infinitely squeezed spin state $W_A(x_A,p_A)
= \delta(x_A-x_0)$ and if the measurement is noiseless.

%%%%%%%%%%%%%%%%%%%%%%%%%%%%%%%%%%%%%%%%%%%%%%%%%%%%%%%%%%%%%%%%%%
%                                                                %
%                           SubSection                           %
%                                                                %
%%%%%%%%%%%%%%%%%%%%%%%%%%%%%%%%%%%%%%%%%%%%%%%%%%%%%%%%%%%%%%%%%%
\subsection{Double-pass schemes}
\label{sec:2pass}

The necessity of preparing the atomic ensemble in a highly squeezed
state or performing a measurement with feedback can be avoided in a
double-pass scheme with two successive, different unitary evolution
\cite{pra74e011802}.  In this scheme, an interaction Hamiltonian
identical to (\ref{eq:1pass}) is applied first.  After a time $t$, the
interaction is suddenly changed to $\hat H_2$, so we have
\begin{gather}
  \label{eq:2pass}
  \hat H_1 = \hat P_A \hat P_L, \quad
  \hat H_2 = \hat X_A \hat X_L ,\\
  \begin{aligned}
  \hat X_A (t+t') &= \hat X_A + t \hat P_L, \\
  \hat P_A (t+t') &= (1-tt') \hat P_A - t' \hat X_L,\\
  \hat X_L (t+t') &= \hat X_L + t \hat P_A, \\
  \hat P_L (t+t') &= (1-tt') \hat P_L - t' \hat X_A.
  \end{aligned}
\end{gather}
If the interaction times are adjusted such that $tt'=1$, we directly
obtain a mapping like (\ref{eq:1pass+feedback}), however, without
measurement and feedback,
\begin{align}
  \hat X_A' &= \hat X_A + t \hat P_L, &
  \hat P_A' &= - \frac1t \hat X_L,
  \nonumber\\\label{eq:2pass'}
  \hat X_L' &= \hat X_L + t \hat P_A, &
  \hat P_L' &= - \frac1t \hat X_A.
\end{align}
Perfect mapping can thus be achieved if the atomic ensemble is
initially prepared in an infinitely squeezed state. In such a scheme
there is no need for measurement and feedback, and the fidelity of the
memory is $F=(\sxa^2/t^2+1)^{-1/2}$ for coherent input light.

Alternatively, as will be shown in the following, the necessity of an
initial atomic squeezing can be avoided by using measurement and
feedback.  Indeed, a measurement of the quadrature $\hat P_L'$ can
project the initial atomic state to a squeezed state---thus
appropriately performing the atomic state preparation after the
interaction~\cite{sci304p270}.  If the measurement of $\hat P_L'$
gives a value $p$, we can replace $\hat X_A$ by $-tp$.  Applying a
position displacement of $tp$ of the atomic position, $\hat
X_A^{\mathrm{mem}} \equiv \hat X_A' + tp$, we obtain an ideal quantum
memory map
\begin{equation}
  \label{eq:2pass+feedback}
  \hat X_A^{\mathrm{mem}} = t \hat P_L, 
  \qquad
  \hat P_A^{\mathrm{mem}} = -\frac 1t \hat X_L.
\end{equation}
Note that an imperfect measurement similar to
(\ref{eq:Pinoisy}) also introduces noise in the position quadrature,
\begin{equation}
  \label{eq:W2pass}
  W^{\mathrm{mem}} (x,p) = \frac1{\sqrt{2\pi\sta^2}}
  \int dx'\, W^{\mathrm{in}}(-tp, x/t + x') e^{-\frac{x'^2}{2\sta^2}}.
\end{equation}
However, light measurement can be performed with far much higher
accuracy than spin squeezing.  For coherent input light, the fidelity
of storage reads $F=(\sta^2+1)^{-1/2}$.

Thus in the double-pass scheme one can get rid of either the
measurement and feedback or the preparation of the atomic
ensemble in a squeezed state.

%%%%%%%%%%%%%%%%%%%%%%%%%%%%%%%%%%%%%%%%%%%%%%%%%%%%%%%%%%%%%%%%%%
%                                                                %
%                           SubSection                           %
%                                                                %
%%%%%%%%%%%%%%%%%%%%%%%%%%%%%%%%%%%%%%%%%%%%%%%%%%%%%%%%%%%%%%%%%%
\subsection{Triple-pass scheme}
\label{sec:3pass}

Finally, we mention that the ideal unitary evolution given in
\sref{sec:pureunitary} can be equivalently achieved in a three-pass
scheme without measurement and with no initial atomic squeezing.  As
it was pointed out in \cite{pra68e022304}, a beam splitter like
interaction Hamiltonian can be simulated by successively applying the
above two kinds of Hamiltonians three times.  In fact if we reapply
$\hat H_1$ for a third time, we obtain
\begin{gather}
  \label{eq:3pass}
  \hat H_1 = \hat P_A \hat P_L, \quad
  \hat H_2 = \hat X_A \hat X_L, \quad
  \hat H_3 = \hat P_A \hat P_L ,\\
  \begin{aligned}
  \hat X_A (t+t'+t'') &= (1-t't'') \hat X_A + [t+t''(1-tt')] \hat P_L, \\
  \hat P_A (t+t'+t'') &= (1-tt') \hat P_A - t' \hat X_L,\\
  \hat X_L (t+t'+t'') &= (1-t't'') \hat X_L + [t+t''(1-tt')] \hat P_A, \\
  \hat P_L (t+t'+t'') &= (1-tt') \hat P_L - t' \hat X_A.
  \end{aligned}
\end{gather}

Setting $t=t'=t''=1$ we arrive at the ideal mapping (\ref{eq:ideal2}).
Let us denote with $\hat U$ the unitary operator describing the time
evolution due to the consecutive actions of $\hat H_1$, $\hat H_2$,
and $\hat H_3$ over periods of time $t=1$, $t'=1$, and $t''=1$,
respectively,
\begin{equation}
  \hat U = e^{-i \hat P_A \hat P_L}
  e^{-i \hat X_A \hat X_L} e^{-i \hat P_A \hat P_L},
\end{equation}
and let us denote with ${\hat U}_{\text{ideal}}$ the unitary evolution
corresponding to the Hamiltonian of the ideal quantum memory
(\ref{eq:Hideal2}) with $\alpha(t)=1$ over a period of $T=\pi/2$,
\begin{equation}
  {\hat U}_{\text{ideal}} = e^{-i\frac\pi2
    (\hat X_A \hat X_L + \hat P_A \hat P_L )}.
\end{equation}
The two unitary operators have the same effect on the quadrature
variables and $\hat U$ and ${\hat U}_{\text{ideal}}$
are in fact identical,
\begin{equation}
  e^{-i \hat P_A \hat P_L}
  e^{-i \hat X_A \hat X_L} 
  e^{-i \hat P_A \hat P_L}
  = e^{-i\frac\pi2 (\hat X_A \hat X_L + \hat P_A \hat P_L )}.
\end{equation}

%%%%%%%%%%%%%%%%%%%%%%%%%%%%%%%%%%%%%%%%%%%%%%%%%%%%%%%%%%%%%%%%%%
%                                                                %
%                           Section                              %
%                                                                %
%%%%%%%%%%%%%%%%%%%%%%%%%%%%%%%%%%%%%%%%%%%%%%%%%%%%%%%%%%%%%%%%%%
\section{Physical systems realizing atomic quantum memory}
\label{sec:memphys}

In this section, we analyze three kinds of configurations that can
serve as collective atomic quantum memory for light.  The first of
them is experimentally carried out by %Julsgaard \textit{et al.}
\citet{nat432p482}.  Then we discuss a scheme based on off-resonant
Raman scattering that realizes the ideal interaction Hamiltonian
(\ref{eq:Hideal2}) directly.  Finally, we show that
electromagnetically induced transparency (EIT) gives rise to an
effective interaction Hamiltonian of type (\ref{eq:Hideal1}).

%%%%%%%%%%%%%%%%%%%%%%%%%%%%%%%%%%%%%%%%%%%%%%%%%%%%%%%%%%%%%%%%%%
%                                                                %
%                           SubSection                           %
%                                                                %
%%%%%%%%%%%%%%%%%%%%%%%%%%%%%%%%%%%%%%%%%%%%%%%%%%%%%%%%%%%%%%%%%%

%%%%%%%%%%%%%%%%%%%%%%%%%%%%%%%%%%%%%%%%%%%%%%%%%%%%%%%%%%%%%%%%%%
\subsection{Quantum memory based on Faraday rotation}
\label{sec:Faraday}

Let us analyze first a scheme in which the light-matter interaction
originates in the paramagnetic Faraday effect~\cite{pra60p4974}:
given an ensemble of atoms with macroscopic magnetic moment and
shined by a linearly polarized light beam propagating in the direction
of the magnetic moment, the plane of light polarization is rotated.

\begin{figure}
  \centering
  \includegraphics[width=\hsize]{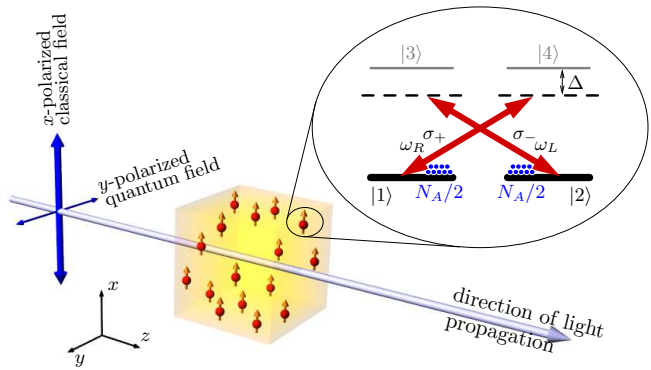}
  \caption{%(color online)
    Quantum memory scheme based on Faraday rotation.  The atoms with
    doubly degenerate ground states are polarized in the polarization
    direction of the strong classical light field.  The inset shows a
    simplified atomic level structure when the quantization axis is
    the direction of light propagation.  The atoms are in an
    equal-weighted coherent superposition of the ground levels.  The
    presence of a weak $y$-polarized light field results in a
    imbalance of the dynamic Stark shifts for the two levels,
    developing a relative phase in the superposition.  The atomic
    spins are, thus, coherently rotated in the $xy$ plane as a
    back-action of light on atoms.}
  \label{fig:css}
\end{figure}

To describe this atom-light interaction, consider an atomic level
structure depicted in the inset of Fig.~\ref{fig:css} where the two
ground levels are off-resonantly coupled to the upper ones by the
right and left circularly polarized electromagnetic field modes of the
same frequencies $\omega_R = \omega_L = \omega$.  In the experiment of
%Julsgaard \textit{et al.} 
\citet{nat432p482}, levels $\ket1$ and
$\ket2$ correspond to the Zeeman sublevels $M_F=\pm4$ of the ground
state $6^2S_{1/2}\,(F=4)$ of cesium, while the light pulses are
detuned to the blue by 700~MHz from the $6^2S_{1/2}\,(F=4) \to
6^2P_{3/2}\,(F=5)$ transition ($\lambda=852\,\mbox{nm}$).

If the detuning $\Delta$ from the atomic transitions is large enough
($g \langle \hat a^\dag \hat a \rangle \ll \Delta$),
we can adiabatically eliminate the off-resonant excited levels so that
the system reduces to an effective two-level atom.  The dynamics is
then governed by the following effective Hamiltonian,
\begin{subequations}
  \begin{eqnarray}
    \label{eq:HFaraday0}
    \hat H_0 &=& \omega 
    (\hat a_R^\dag \hat a_R + \hat a_L^\dag \hat a_L +1)
    \nonumber\\&&
    + E_0 \sum_{i=1}^{N_A} (\ket1_{ii}\bra1 + \ket2_{ii}\bra2),\\
    \label{eq:HFaraday1}
    \!\!
    \hat H_\textint &=& \sum_{i=1}^{N_A} \frac{g^2}{\Delta} 
    \big[\hat a_R^\dag \hat a_R \ket1_{ii}\bra1 
      + \hat a_L^\dag \hat a_L \ket2_{ii}\bra2 \big], 
  \end{eqnarray}
\end{subequations}
where $N_A$ is the number of atoms and the electric dipole coupling
constant for the single-photon transitions is
\begin{equation}
  g = \sqrt{\frac{\omega}{2V}} 
  \big| \bra1 \hat{\mathbf d} \boldsymbol\epsilon_R \ket4 \big|
  = \sqrt{\frac{\omega}{2V}} 
  \big| \bra2 \hat{\mathbf d} \boldsymbol\epsilon_L \ket3 \big|.
\end{equation}
Note that \eqref{eq:HFaraday1} is nothing more than the dynamic Stark
shift (light shift) of the energy of the lower levels caused by
virtual transitions to the off-resonant upper levels.

Let us introduce the quantum mechanical Stokes parameters to describe
the polarization state of light,
\begin{eqnarray}
  \nonumber%\label{eq:defSx}
  \hat S_x &\equiv& (\hat a_R^\dag \hat a_L + \hat a_L^\dag \hat
  a_R)/2 = (\hat a_x^\dag \hat a_x - \hat a_y^\dag \hat a_y)/2, 
  \\
  \nonumber%\label{eq:defSy}
  \hat S_y &\equiv& (\hat a_R^\dag \hat a_L - \hat a_L^\dag \hat
  a_R)/(2i) = (\hat a_x^\dag \hat a_y + \hat a_y^\dag \hat a_x)/2, 
  \\
  \nonumber%\label{eq:defSz}
  \hat S_z &\equiv& (\hat a_R^\dag \hat a_R - \hat a_L^\dag \hat
  a_L)/2 = (\hat a_x^\dag \hat a_y - \hat a_y^\dag \hat a_x)/(2i),
  \\
  \label{eq:defStokes}
  \hat N_S &\equiv& (\hat a_R^\dag \hat a_R + \hat a_L^\dag \hat
  a_L)/2 = (\hat a_x^\dag \hat a_x + \hat a_y^\dag \hat a_y)/2.
\end{eqnarray}
$\hat S_x$ is the photon number (intensity) difference of the $x$ and
$y$ linearly polarized light components, $\hat S_y$ is that of the
diagonally polarized ones, $\hat S_z$ corresponds to the difference in
the right and left circularly polarized components, while $\hat N_S$
is half of the total photon number in the two modes.
It is easy to see that they satisfy the standard angular momentum
commutation relations $[\hat S_\alpha, \hat S_\beta] =i
\epsilon_{\alpha \beta \gamma} \hat S_\gamma$ and $[\hat N_S, \hat
{\mathbf S}] = \mathbf 0$.  The annihilation operators $\hat a_x =
(\hat a_R + \hat a_L)/\sqrt2$ and $\hat a_y = (\hat a_R - \hat
a_L)/(i\sqrt2)$ correspond to the linearly polarized components of the
light beam.  Note that $\hat {\mathbf S}$ is not a vector operator,
i.e., it does not transform as a vector under rotations.

For the atomic magnetic moment, we can use Schwinger's
representation and introduce the collective atomic
quasi-spin variables,
\begin{eqnarray}
  \nonumber%\label{eq:defFx}
  \hat \sigma_x &\equiv& \sum_i 
      \big( \ket2_{ii}\bra1 + \ket1_{ii}\bra2 \big) /2,
  \\
  \nonumber%\label{eq:defFy}
  \hat \sigma_y &\equiv& \sum_i 
      \big( \ket2_{ii}\bra1 - \ket1_{ii}\bra2 \big) /2i,
  \\
  \nonumber%\label{eq:defFz}
  \hat \sigma_z &\equiv& \sum_i 
      \big( \ket2_{ii}\bra2 - \ket1_{ii}\bra1 \big) /2,
  \\
  \label{eq:defquasispin}
  \hat N_\sigma &\equiv& \sum_i 
      \big( \ket2_{ii}\bra2 + \ket1_{ii}\bra1 \big) /2,
\end{eqnarray}
where $2\hat N_\sigma$ is equal to the number of atoms in the subspace
spanned by $\ket1$ and $\ket2$, i.e., $\hat N_\sigma$ counts the atoms
contributing to the phenomenon.  Note that unless we have a real
spin system, the quasi-spin vector is not a vector operator either
because $[\hat \sigma_\alpha, \hat F_\beta] \ne i \epsilon_{\alpha
  \beta \gamma} \hat \sigma_\gamma$ with $\hat{\mathbf F}$ being the
total angular momentum.  By definition, the components
satisfy the commutation relations
\begin{equation}
  \label{eq:sigmacommute}
  [\hat \sigma_\alpha, \hat \sigma_\beta] 
  =i \epsilon_{\alpha \beta \gamma} \hat \sigma_\gamma.
\end{equation}

We are now ready to express the effective interaction Hamiltonian
(\ref{eq:HFaraday1}) in terms of the Stokes and quasi-spin vectors,
\begin{subequations}\label{eq:HFaraday2ab}
  \begin{eqnarray}
    \hat H_0 &=& \omega (2\hat N_S + 1) + 2E_0 \hat N_\sigma, \\
    \label{eq:HFaraday2}
    \hat H_\textint &=& \frac{2g^2}{\Delta} 
    \left(\hat N_S \hat N_\sigma - \hat S_z \hat\sigma_z \right). 
  \end{eqnarray}
\end{subequations}
The operators $\hat N_S$, $\hat N_\sigma$, and $\hat N_S \hat
N_\sigma$ all commute with both $\hat{\mathbf S}$ and
$\hat{\boldsymbol\sigma}$ (and thus, also with $\hat S_z
\hat\sigma_z$) and they have nothing to do with the dynamics.
Therefore, we can safely omit them and write
\begin{equation}
  \label{eq:HFaraday3}
  \hat H_0 = \text{const.}, 
  \qquad
  \hat H_\textint = - \frac{2g^2}{\Delta} 
  \hat S_z \hat\sigma_z. 
\end{equation}
The remaining term explains paramagnetic Faraday
rotation~\cite{pra60p4974}.  If the $z$ component of the collective
atomic quasi-spin has a macroscopic expectation value, it results in
an interaction Hamiltonian proportional to $\hat S_z$ which in turn
introduces rotation of the Stokes vector along the $z$ axis---thus
turning the linear polarization in the $xy$ plane.  Reversely, as a
back-action of light on atoms, if the $z$ component of the Stokes
vector has a macroscopic expectation value, it will rotate coherently
each atomic spin along the $z$ axis.

Now we show---following Ref.~\cite{nat432p482}---that \eref{eq:HFaraday3}
gives rise to a Hamiltonian of the form (\ref{eq:1pass}).
Consider an atomic ensemble in which all atoms are initially prepared
in the superposition state $(\ket1+\ket2)/\sqrt2$.  In this coherent
spin state (CSS), the $x$ component of the quasi-spin vector has a
macroscopic expectation value $\langle \hat\sigma_x \rangle_0 =
N_A/2$.  This state is actually a quasi-spin-$\frac{N_A}2$ state,
i.e., an eigenstate of $\hat{\boldsymbol\sigma}^2$ with eigenvalue
$\frac{N_A}2 (\frac{N_A}2+1)$.  Since the Hamiltonian
\eqref{eq:HFaraday2ab} commutes with  $\hat{\boldsymbol\sigma}^2$, the
atomic state will always remain a quasi-spin-$\frac{N_A}2$ state.
Moreover, as long as the interaction introduces small perturbation to
$\hat{\boldsymbol\sigma}$, the $y$ and $z$ components of
$\hat{\boldsymbol\sigma}$ stay small with respect to the $x$
component.  Therefore, we can use the Holstein-Primakoff
approximation~\cite{pr58p1098} and introduce the position and
momentum-like quadrature operators
\begin{equation}
  \label{eq:defXAPAFaraday}
  \hat X_A \equiv \hat\sigma_y / \sqrt{\langle \hat\sigma_x \rangle_0},
  \qquad
  \hat P_A \equiv \hat\sigma_z / \sqrt{\langle \hat\sigma_x \rangle_0}.
\end{equation}
As long as the number of atomic excitations $\hat n_A = \frac12 (\hat
X_A^2 + \hat P_A^2 -1)$ is small compared to the number of atoms, the
deviation of the quasi-spin vector from the CSS stays in the tangent
plane of the spin sphere.  Then \eref{eq:sigmacommute} ensures that we
have the approximately correct commutation relation
\begin{equation}
  \label{eq:XAPAcommute}
  [\hat X_A, \hat P_A] = i - \frac{2i}{N_A} \hat n_A \approx i.
\end{equation}

In the experiment of~\cite{nat432p482}, the quantum information is
represented in the $y$-polarized weak signal beam propagating in the
$z$ direction and it is mixed with the copropagating $x$-polarized
strong classical control field having coherent amplitude $\alpha$
(Fig.~\ref{fig:css}).  Writing the c-number $\alpha$ in place of $\hat
a_x$, we find that the $x$ component of the Stokes vector
(\ref{eq:defStokes}) has a macroscopic expectation value, $\langle\hat
S_x\rangle = |\alpha|^2/2$ in first order.  Therefore, we can
introduce the light quadrature variables as
\begin{eqnarray}
  \nonumber
  \hat X_L &\equiv& \hat S_y / \sqrt{\langle \hat S_x \rangle_0}
  = \big(e^{-i\varphi}\hat a_y + e^{i\varphi}\hat a_y^\dag
  \big)/\sqrt2,
  \\\label{eq:defXLPLFaraday}
  \hat P_L &\equiv& \hat S_z / \sqrt{\langle \hat S_x \rangle_0}
  = \big(e^{-i\varphi}\hat a_y - e^{i\varphi}\hat a_y^\dag
  \big)/i\sqrt2,
\end{eqnarray}
where $\varphi=\arg(\alpha)$ is the complex phase of $\alpha$.

In summary we express the Stokes and
quasi-spin vectors in terms of the quadrature operators
[cf.~Eqs.~(\ref{eq:defStokes}) and~(\ref{eq:defquasispin})],
\begin{align}
  \nonumber
  \hat S_x &= |\alpha|^2/2 - \hat n_L,&       \hat \sigma_x &= N_A/2 - \hat n_A,\\
  \nonumber
  \hat S_y &= \hat X_L \sqrt{|\alpha|^2/2},&  \hat \sigma_y &= \hat X_A \sqrt{N_A/2},\\
  \nonumber
  \hat S_z &= \hat P_L \sqrt{|\alpha|^2/2},&  \hat \sigma_z &= \hat P_A \sqrt{N_A/2},\\
  \hat N_S &= |\alpha|^2/2,&                  \hat N_\sigma &= N_A/2.
\end{align}
Note that these operators exactly satisfy the angular momentum-like
commutation relations $[\hat N_\sigma, \hat{\boldsymbol\sigma}] =
\mathbf 0$, $[\hat \sigma_\alpha, \hat \sigma_\beta] =i
\epsilon_{\alpha \beta \gamma} \hat \sigma_\gamma$, $[\hat N_S, \hat
{\mathbf S}] = \mathbf 0$, and $[\hat S_\alpha, \hat S_\beta] =i
\epsilon_{\alpha \beta \gamma} \hat S_\gamma$, despite the fact that
we used an oscillator approximation for the atomic system and a
classical model for the $x$-polarized light.

Putting (\ref{eq:defXAPAFaraday}) and (\ref{eq:defXLPLFaraday})
into (\ref{eq:HFaraday3}) we arrive at the interaction Hamiltonian in
terms of the atomic and light quadrature operators,
\begin{equation}
  \label{eq:HFaraday4}
  \hat H_\textint = - \frac{g^2 |\alpha| \sqrt{N_A}}\Delta
  \hat P_L \hat P_A.
\end{equation}
We recover the non-ideal interaction Hamiltonian \eqref{eq:1pass}.  As
a consequence, the atomic position quadrature $\hat X_A$ is displaced
by an amount proportional to $\hat P_L$ [c.f.~\eref{eq:1passevol}],
thus rotating the quasi-spin vector towards the $y$ axis.

%%%%%%%%%%%%%%%%%%%%%%%%%%%%%%%%%%%%%%%%%%%%%%%%%%%%%%%%%%%%%%%%%%
%                                                                %
%                           SubSection                           %
%                                                                %
%%%%%%%%%%%%%%%%%%%%%%%%%%%%%%%%%%%%%%%%%%%%%%%%%%%%%%%%%%%%%%%%%%
%%%%%%%%%%%%%%%%%%%%%%%%%%%%%%%%%%%%%%%%%%%%%%%%%%%%%%%%%%%%%%%%%%
\subsection{Off-resonant Raman scheme}
\label{sec:Raman}

Now we discuss another physical system which directly realizes the
ideal interaction Hamiltonian (\ref{eq:Hideal2}).  Our quantum memory
scheme is based on off-resonant Raman scattering by $\Lambda$-type
atoms \cite{pra62e033809}.

\begin{figure}
  \centering
  \includegraphics[width=\hsize]{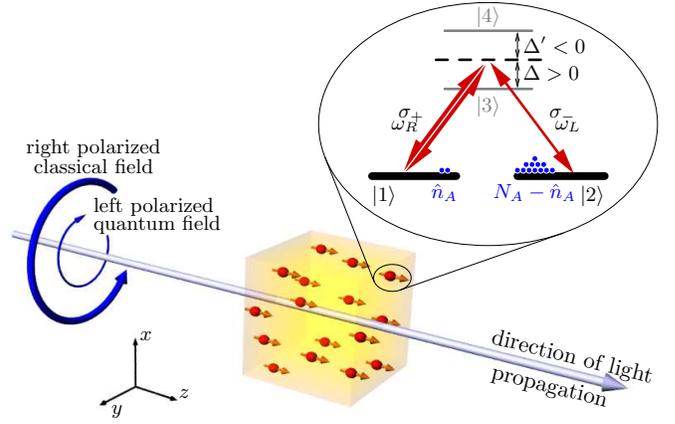}
  \caption{%(color online)
    Quantum memory scheme based on off-resonant Raman scattering.  The
    inset shows the $\Lambda$-type atomic configuration.  The
    degenerate ground states $\ket1$ and $\ket2$ are off-resonantly
    coupled to the intermediate levels $\ket3$ and $\ket4$ via right
    and left circularly polarized monochromatic light beams.  Light
    shifts of the lower levels are cancelled by tuning the laser
    fields right between the two upper levels.  The classical control
    field is the right polarized one.  Only level $\ket2$ is
    macroscopically populated, which corresponds to a quasi-spin
    polarization in the direction of light propagation.  Absorption of
    a signal photon results a collective atomic excitation in level
    $\ket1$.}
  \label{fig:css2}
\end{figure}

Consider an ensemble of $N_A$ atoms depicted in Fig.~\ref{fig:css2}
that is shined by a monochromatic light beam consisting of
copropagating phase-locked right and left circularly polarized
components.  Levels $\ket1$ and $\ket2$ can be, for example, the
$m_F=\pm1$ Zeeman sublevels of the ${}^2S_{1/2}\,(F=1)$ ground state
of sodium or rubidium (with nuclear spin $I=3/2$), while levels
$\ket3$ and $\ket4$ are the $m_F=0$ Zeeman sublevels of the hyperfine
levels ${}^2P_{1/2}\,(F=1)$ and ${}^2P_{1/2}\,(F=2)$.  Although the
Zeeman sublevels $M_F=\pm2$ of the uppermost level
${}^2P_{1/2}\,(F=2)$ also contributes to the light shifts, the
principle of the model is not changed.

%%%%%%%%%%%%%%%%%%%%%%%%%%%%%%%%%%%%%%%%%%%%%%%%%%%%%%%%%%%%%%%%%
%%%%%%%%%%%%%%%%%%%%%%%%%%%%%%%%%%%%%%%%%%%%%%%%%%%%%%%%%%%%%%%%

If the atomic transitions are far off-resonant, the four-level atom is
reduced to an effective two-level one.  After adiabatic elimination of
the upper levels, the dynamics is governed by an effective interaction
Hamiltonian that consists of dynamic Stark shifts and two-photon
processes between the ground states,
\begin{multline}
  \label{eq:HLambda1}
    \hat H_\textint = \sum_{i=1}^{N_A} \bigg\{
    \bigg(\frac{|g_{3R}^{{i}}|^2}{\Delta}
    + \frac{|g_{4R}^{{i}}|^2}{\Delta'}\bigg)
    \hat a_R^\dag \hat a_R \ket1_{ii}\bra1
    \\
    +\bigg(\frac{|g_{3L}^{{i}}|^2}{\Delta}
    + \frac{|g_{4L}^{{i}}|^2}{\Delta'}\bigg)
    \hat a_L^\dag \hat a_L \ket2_{ii}\bra2
    \\ +\bigg[ 
    \bigg(\frac{g_{3R}^{{i}} g_{3L}^{{i}} \strut^*}{\Delta}
    + \frac{g_{4R}^{{i}} g_{4L}^{{i}} \strut^*}{\Delta'}\bigg)
    \hat a_L^\dag \hat a_R \ket2_{ii}\bra1 + \mathrm{H.c.}
    \bigg] \bigg\},
\end{multline}
where the electric dipole coupling constants for the single-photon
transitions are
\begin{equation}
  g_{3R}^{{i}} = \sqrt{\frac{\omega_R}{2\hbar V}} \bra1 
  \hat{\mathbf d} \cdot \boldsymbol\epsilon_R \ket3 
  e^{i(\mathbf k_R \mathbf r_i + \phi_R)},
\end{equation}
and similarly for $g_{3L}^{{i}}$, $g_{4R}^{{i}}$, and $g_{4L}^{{i}}$,
with $\mathbf r_i$ being the position of the $i$th atom.  For symmetry
reasons, we have $|g_{3R}^{{i}}| = |g_{3L}^{{i}}| \equiv g$ and
$|g_{4R}^{{i}}| = |g_{4L}^{{i}}| \equiv g'$.  Furthermore, since the
frequencies and propagation directions of the two polarized light
beams coincide, we can choose the relative phase $\phi_R-\phi_L$ of
the mode functions of
the light modes so that $g_{3R}^{{i}} g_{3L}^{{i}} \strut^* = g^2$ is
real and positive.  However, in our example of sodium or rubidium,
$g_{4R}^{{i}}$ and $g_{4L}^{{i}}$ have opposite signs for this choice
of phases, so $g_{4R}^{{i}} g_{4L}^{{i}} \strut^* = -g'^2$.

We can introduce the photonic Stokes vector $ {\hat{\mathbf S}}$
(\ref{eq:defStokes}) and the collective atomic quasi-spin vector
${\hat {\mathbf \sigma}}$ (\ref{eq:defquasispin}).  If the ground
states are $F=1$ hyperfine sublevels, the components of the quasi-spin
expressed by the total angular momentum $\hat{\mathbf F}$ are
\begin{eqnarray}
  \nonumber
  \hat \sigma_x &=& \sum_i \frac12
      \big(\hat F_x^{(i)}\strut^2 
          - \hat F_y^{(i)}\strut^2 \big),
  \\\nonumber
  \hat \sigma_y &=& \sum_i \frac12
      \big(\hat F_x^{(i)} \hat F_y^{(i)}
         + \hat F_y^{(i)} \hat F_x^{(i)}\big),
  \\\label{eq:defsigmaraman}
  \hat \sigma_z  &=& \hat F_z /2. 
\end{eqnarray}

We are now ready to simplify the interaction Hamiltonian
(\ref{eq:HLambda1}).  We realize that the first two terms therein are
proportional to $\hat N_S \hat N_\sigma - \hat S_z \hat \sigma_z$ and
they are responsible for paramagnetic Faraday rotation.  Since $\hat
N_S \hat N_\sigma$ commutes with both $\hat{\mathbf S}$ and
$\hat{\mathbf \sigma}$ and has nothing to do with the dynamics, we
will omit it in the following.  The last term in (\ref{eq:HLambda1})
is proportional to $\hat S_x \hat \sigma_x + \hat S_y \hat \sigma_y$.
All in all, we have
\begin{multline}
  \hat H_\textint = 
  -2\left(\frac{g^2}{\Delta} + \frac{g'^2}{\Delta'}\right)
     \hat S_z \hat \sigma_z
  \\
  +2\left(\frac{g^2}{\Delta} - \frac{g'^2}{\Delta'}\right)
     \left(\hat S_x \hat \sigma_x + \hat S_y \hat \sigma_y\right).
  \label{eq:HLambda3}
\end{multline}
We note that the effective Hamiltonian is essentially of the same form
even for more general configurations involving multiple atomic
levels~\cite{pra71e032348}.  The meaning of the first term is
explained in Sec.~\ref{sec:Faraday}.  By tuning the laser fields right
between the two upper levels so that $g'^2/\Delta' = -g^2/\Delta$, we
can cancel the first term.  The remaining term then describes the
two-photon processes of the atoms making transitions between the two
ground states.  However, this term can also be interpreted as Faraday
rotation.  Suppose, for example, that the $y$ component of the Stokes
vector has a macroscopic expectation value, that is, the atomic
ensemble is irradiated by a single $45^\circ$ linear polarized
classical beam.  Then the interaction Hamiltonian (\ref{eq:HLambda3})
reduces to a term proportional to $\hat\sigma_y$ corresponding to a
coherent rotation of the quasi-spin along the $y$ axis.  Indeed, if we
choose $y$ as the quantization axis, such a diagonally polarized light
will induce virtual atomic transitions from the superposition state
$(\ket1+i\ket2)/\sqrt2$ to the off-resonant upper levels, thus
shifting the energy level of this superposition state with respect to
the orthogonal state $(\ket1-i\ket2)/\sqrt2$.  As a consequence, all
the atomic quasi-spins are rotated along the $y$-axis.  Appropriately
choosing the polarization and detuning of a single classical light
field, one can realize rotation of the quasi-spin vector along
\emph{arbitrary} axis.

To use the present configuration as a quantum memory, the
polarizations of the control and signal fields and the initial atomic
state should be chosen differently
from that of Sec.~\ref{sec:Faraday}.  Let us represent the quantum
information in the weak left circularly polarized light beam (signal) and let
the right circularly polarized light be the strong classical field (control)
with coherent amplitude $\alpha$ (though not too strong so that it
remains off-resonant, $|\alpha|\ll|\Delta/g|$, $|\Delta'/g'|$).  We
can write the c-number $\alpha$ instead of the right circularly
polarized annihilation operator $\hat a_R$ and we find that the z
component of the Stokes vector is a classical variable $\langle \hat
S_z \rangle_0 = |\alpha|^2/2$.  This enables us to introduce the light
quadrature variables as
\begin{eqnarray}
  \nonumber%\label{eq:EITXL}
  \hat X_L &\equiv& \hat S_x / {\sqrt{|\alpha|^2/2}}
  = ({e^{-i\varphi} \hat a_L + e^{i\varphi} \hat a_L^\dag})/{\sqrt2},\\
  \label{eq:EITPL}
  \hat P_L &\equiv& {\hat S_y}/{\sqrt{|\alpha|^2/2}}
  = ({e^{-i\varphi} \hat a_L - e^{i\varphi} \hat a_L^\dag})/{i\sqrt2},
\end{eqnarray}
where $\varphi=\arg(\alpha)$ is the phase of $\alpha$.  Similarly we
can introduce quadratures for the atomic ensemble if the collective
atomic state stays close to the coherent spin state (CSS) in which all
atoms are in level~$\ket2$.  The $z$ component of the quasi-spin has a
macroscopic expectation value, $\langle \hat \sigma_z \rangle_0 =
N_A/2$ in zeroth order, and we can approximate it with $\hat \sigma_z
= N_A/2 - \hat n_A$.  Then we can write the atomic quadrature
variables as
\begin{equation}
  \label{eq:defXAPAEIT}
  \hat X_A \equiv \frac{\hat\sigma_x}{\sqrt{N_A/2}},
  \qquad
  \hat P_A \equiv \frac{\hat\sigma_y}{\sqrt{N_A/2}}.
\end{equation}

In order to keep the correct angular momentum commutation
relations of the Stokes and quasi-spin vectors, we write 
\begin{align}
  \nonumber
  \hat S_x &= \hat X_L \sqrt{|\alpha|^2/2},&   \hat \sigma_x &= \hat X_A \sqrt{N_A/2},\\
  \nonumber
  \hat S_y &= \hat P_L \sqrt{|\alpha|^2/2},&  \hat \sigma_y &= \hat P_A \sqrt{N_A/2},\\
  \nonumber
  \hat S_z &= |\alpha|^2/2 - \hat n_L,&       \hat \sigma_z &= N_A/2 - \hat n_A,\\
  \hat N_S &= |\alpha|^2/2,&                  \hat N_\sigma &= N_A/2.
\end{align}
Note that the number of atomic excitations $\hat n_A = \frac12(\hat
X_A^2 + \hat P_A^2 -1)$ equals to the population of level $\ket1$,
$\hat n_A = \sum_i \ket1_{ii}\bra1$.  Putting all together while
canceling the term $\hat S_z \hat\sigma_z$ in \eqref{eq:HLambda3}, we
obtain the interaction Hamiltonian
\begin{equation}
  \label{eq:HLambda4}
  \hat H_\textint 
  = \frac{2g^2|\alpha|\sqrt{N_A}}{\Delta}
  (\hat X_L \hat X_A + \hat P_L \hat P_A).
\end{equation}

The Hamiltonian (\ref{eq:HLambda4}) results in an oscillation of the
excitations between modes $A$ and $L$.  Indeed, with the creation and
annihilation operators of systems $A$ and $L$ we recognize the beam
splitter Hamiltonian
\begin{equation}
  \label{eq:xx+pp=aa+aa}
  \hat X_L \hat X_A + \hat P_L \hat P_A
  =\hat a_L^\dag \hat a_A + \hat a_A^\dag \hat a_L.
\end{equation}
Absorption of a left polarized photon makes an atomic transition from
level $\ket2$ to level $\ket1$, and reversely, emission of such a
photon causes a decrease in the population $\hat n_A$ of level
$\ket1$.

The state of the two systems can be exchanged completely by
appropriately adjusting the amplitude $\alpha(t)$ of the classical
right circularly polarized control field, so that we have a $\pi/2$ pulse,
\begin{equation}
  \label{eq:intpi2}
  \frac{2g^2\sqrt{N_A}}{\Delta}
  \int |\alpha(t)|\,dt = \frac\pi2.
\end{equation}
This leads to the ideal quantum memory mapping (\ref{eq:ideal2}).

%%%%%%%%%%%%%%%%%%%%%%%%%%%%%%%%%%%%%%%%%%%%%%%%%%%%%%%%%%%%%%%%%%
%                                                                %
%                           SubSection                           %
%                                                                %
%%%%%%%%%%%%%%%%%%%%%%%%%%%%%%%%%%%%%%%%%%%%%%%%%%%%%%%%%%%%%%%%%%
%%%%%%%%%%%%%%%%%%%%%%%%%%%%%%%%%%%%%%%%%%%%%%%%%%%%%%%%%%%%%%%%%%
\subsection{Resonant EIT scheme}
\label{sec:EIT}

\begin{figure}
  \centering
    \includegraphics{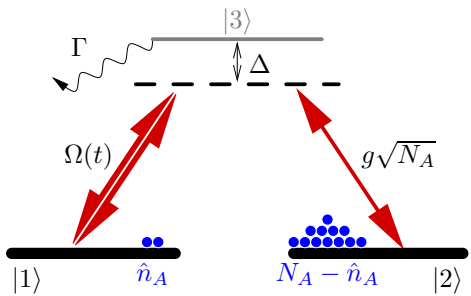}
    \caption{%(color online)
      $\Lambda$-type atomic configuration in the EIT scheme.  Levels
      $\ket1$ and $\ket3$ are coupled by the classical field of Rabi
      frequency $\Omega(t)$ which is controlled externally.  Levels
      $\ket2$ and $\ket3$ are coupled by the weak signal field.
      Initially, only level $\ket2$ is populated.}
  \label{fig:eit-atom}
\end{figure}

In quantum memories based on EIT \cite{prl84p5094}, an intense
classical radiation field (control) and a weak quantum field to be
stored (signal) are adjusted on or near resonance with the transitions
of the $\Lambda$-type atoms (see Fig.~\ref{fig:eit-atom}).  In this
subsection we show how to adiabatically eliminate the resonant excited
level and we derive a beam splitter-like effective Hamiltonian between
the signal light field and the collective coherences of the lower
atomic levels.  In order to simplify the discussion, we will restrict
ourselves to single-mode radiation fields for the control and signal
with exact two-photon resonance and standing polariton wave.  The
Hamiltonian in the rotating frame then reads
\begin{equation}
    \label{eq:Heit1}
  \hat H = \sum_{i=1}^{N_A} \Big[ 
  -\Delta \ket3_{ii}\bra3 \\
  + \bigl(g_i \hat a_L \ket3_{ii}\bra2
  + \Omega_i \ket3_{ii}\bra1 + \mathrm{H.c.} \bigr)\Big],
\end{equation}
where $\hat a_L$ is the bosonic operator of the signal field, $\Delta$
is the one-photon detuning for both transitions.  Furthermore, we
disregard atomic motion and assume that the coupling constants $g_i$
are real and the same for all atoms, and so are the Rabi frequencies
$\Omega_i$.  This allows us to introduce collective spin operators
\begin{equation}
  \label{eq:sigmaab}
  \hat \sigma_{ab} \equiv \sum_i \ket a_{ii}\bra b
  \qquad (a,b=1,2,3).
\end{equation}

%%%%%%%%%%%%%%%%%%%%%%%%%%%%%%%%%%%%%%%%%%%%%%%%%%%%%%%%%%%%%%%%
%%%%%%%%%%%%%%%%%%%%%%%%%%%%%%%%%%%%%%%%%%%%%%%%%%%%%%%%%%%%%%%%%%

Due to the symmetry of the Hamiltonian (\ref{eq:Heit1}), only the
totally symmetric Dicke states are coupled to the light fields.  The
totally symmetric state containing $n$ atoms on level $\ket1$, $m$
atoms on level $\ket2$, and $l$ on level $\ket3$ is defined as
\begin{multline}
  \label{eq:defdicke}
  \ket{n_1, m_2, l_3}\equiv \big[n!\,m!\,l!\,(n+m+l)!\big]^{-1/2}
  \\\times
  \sum \ket{ 1\colon i_1,\ldots,i_n;\, 
    2\colon j_1,\ldots,j_m;\,  3\colon k_1,\ldots,k_l},
\end{multline}
where the summation is over the $N_A=n+m+l$ mutually different variables $i_1$,~\ldots,
$i_n$, $j_1$, \ldots, $j_m$, $k_1$, \ldots,~$k_l$ going from 1 to
$N_A$.  The ket vector at the
right of (\ref{eq:defdicke}) represents the atomic product state in
which atoms indexed by $i_1$,~\ldots, $i_n$ are in state $\ket1$, etc.
It is easy to verify that the spin-flip operators (\ref{eq:sigmaab})
keep the symmetry of these states,
\begin{eqnarray}
  \nonumber
  \hat \sigma_{ab} \ket{n_a, m_b, l_c} &=&
  \sqrt{(n+1)m} \ket{(n+1)_a, (m-1)_b, l_c},\\
  \hat \sigma_{aa} \ket{n_a, m_b, l_c} &=&
  n \ket{n_a, m_b, l_c},
\end{eqnarray}
with $\{a,b,c\}=\{1,2,3\}$.  We can see that for each value of
$n=0$,~\ldots, $N_A$, the set
\begin{multline}
  \label{eq:eitSn}
  \mathcal S_n \equiv \Span \big\{ \ket{ (k-l)_1, (N_A-k)_2, l_3}
  \otimes \ket{n-k}_L \big| \\ k=0,\ldots, \min(n,N_A) \mbox{ and }
  l=0,\ldots,k \big\}
\end{multline}
is an invariant subspace of the Hamiltonian \eqref{eq:Heit1} so that
\eqref{eq:Heit1} is block diagonal.  (The subscript $L$ means that the
corresponding Fock state refers to the signal light mode.)

Now we make use of the fact that only the atomic level $\ket2$ is
macroscopically populated, so $\hat\sigma_{22}$ can be substituted by
the c-number $N_A$.  We observe that only the coherences
$\hat\sigma_{21}$ and $\hat\sigma_{23}$ are important, as we can
express the other spin operators as $\hat\sigma_{11} = \hat\sigma_{12}
\hat\sigma_{21} /N_A$, $\hat\sigma_{33} = \hat\sigma_{32}
\hat\sigma_{23} /N_A$, and $\hat\sigma_{13} = \hat\sigma_{12}
\hat\sigma_{23} /N_A$.  Compared to the single atomic oscillator in
the off-resonant Raman process investigated in Sec.~\ref{sec:Raman},
now we have two atomic oscillator modes with annihilation operators
$\hat\sigma_{21} /\sqrt{N_A}$ and $\hat\sigma_{23} /\sqrt{N_A}$,
respectively.  Within the subspace ${\mathcal S} \equiv
\bigoplus_{n\ll N_A} {\mathcal S_n}$ of photonic and totally symmetric
atomic states with arbitrary but small number of excitations $n$, the
Hamiltonian can be written in the matrix form %
\footnote{%
  We note that $\mathcal S^\perp$, the complementer subspace of
  $\mathcal S$, consists of non-symmetric excitations that are
  uncoupled to the polaritons and do not influence the behavior of the
  quantum memory \citep{pra72e022327}.%
}%
\begin{equation}
  \label{eq:matrix-Ham}
  \hat H = \hat{\mathbf z}^\dag \mathbf H \hat{\mathbf z},
  \quad
  \mathbf H \equiv 
  \begin{pmatrix}
    0 & 0 & g\sqrt{N_A} \\
    0 & 0 & \Omega \\
    g\sqrt{N_A} & \Omega & -\Delta
  \end{pmatrix}.
\end{equation}
Here we have introduced the vector notation $\hat{\mathbf z}^\dag
\equiv ( \hat a_L^\dagger,\, {\hat\sigma_{12}}/{\sqrt{N_A}},\,
{\hat\sigma_{32}}/{\sqrt{N_A}} )$ for the creation operators of the
photonic mode and the two atomic oscillator modes.  Note that in the
limit of small atomic excitation, components of $\hat{\mathbf z}$ and
$\hat{\mathbf z}^\dag$ approximately satisfy bosonic commutation
relations. The matrix $\mathbf H$ can be brought to a block diagonal
form.  The corresponding transformation matrix $\mathbf R$ defines the
quantum field variables (annihilation operators) of the so-called
dark- and bright-state polaritons \cite{progopt35p257, prl84p4232,
  prl84p5094},
\begin{equation}
  \hat{\mathbf z}' \equiv
  \begin{pmatrix}
    \hat\Psi\\\hat\Phi\\\hat\Xi
  \end{pmatrix} 
  \equiv \mathbf R \hat{\mathbf z},
  \quad \mathbf R = 
  \begin{pmatrix}
    \cos\theta & -\sin\theta & 0\\
    \sin\theta & \cos\theta & 0 \\
    0 & 0 & 1
  \end{pmatrix}
  ,
  \label{eq:defpolariton}
\end{equation}
where $\hat\Psi$, $\hat\Phi$, and $\hat\Xi$ stand for the
dark-polariton, the bright-polariton, and excited-state modes,
respectively. The mixing angle~$\theta$ is defined as
\begin{equation}
  \label{eq:mixingangle}
  \tan\theta\equiv\frac{g\sqrt{N_A}}\Omega.
\end{equation}
Note that in the limit of small atomic excitations, the polariton
operators also satisfy the bosonic commutation relations,
\begin{align}
  \nonumber
  [\hat\Psi, \hat\Psi^\dag]
  & \approx 
%  &= 1 - (2\hat\sigma_{11}+\hat\sigma_{33}) \sin^2\theta /N_A
%  ,\\\nonumber
  [\hat\Phi, \hat\Phi^\dag]
  \approx
%  &= 1 - (2\hat\sigma_{11}+\hat\sigma_{33}) \cos^2\theta /N_A
%  ,\\\nonumber
  [\hat\Xi, \hat\Xi^\dag] 
  \approx 1
%  &= 1 - (\hat\sigma_{11}+2\hat\sigma_{33}) /N_A
  ,\\\nonumber
%% Cross terms.
  [\hat\Psi, \hat\Phi^\dag] 
  & \approx
%  &= 0 + (2\hat\sigma_{11}+\hat\sigma_{33}) \sin\theta \cos\theta /N_A
%  ,\\\nonumber
  [\hat\Psi, \hat\Xi^\dag]
  \approx
%  &= 0 + \hat\sigma_{31} \sin\theta /N_A
%  ,\\\nonumber
  [\hat\Phi, \hat\Xi^\dag]
  \approx 0
%  &= 0 - \hat\sigma_{31} \cos\theta /N_A
  ,\\
%% Remainings.
  [\hat\Psi, \hat\Phi] &=
  [\hat\Psi, \hat\Xi] =
  [\hat\Phi, \hat\Xi] = 0.
\end{align}

Actually, the rotation \eqref{eq:defpolariton} corresponds to
switching from Heisenberg picture to a rotating axes representation.
The generator of the unitary transformation is the Hermitian operator
\begin{equation}
  \label{eq:defK}
  \hat K \equiv i \big(\hat\Psi^\dag \hat\Phi
  - \hat\Phi^\dag \hat\Psi \big)
  = i (\hat a_L^\dag \hat\sigma_{21} 
  - \hat a_L \hat\sigma_{12} )/{\textstyle\sqrt{N_A}}.
\end{equation}
Operators in the rotating axes representation are obtained from those
in the Heisenberg picture in the following way,
\begin{equation}
  \hat{\mathbf z}' \equiv e^{-i\theta \hat K}
  \hat{\mathbf z} e^{i\theta \hat K}.  
\end{equation}

The equations of motion for the rotating polariton variables are
\begin{equation}
  \frac d{dt} \hat{\mathbf z}' = 
  \frac d{dt} \big( \mathbf R \hat{\mathbf z} \big) = 
  i [\hat H, \mathbf R \hat{\mathbf z}] 
  + \left(\frac d{dt} \mathbf R \right) \hat{\mathbf z}
  = i [\hat{\tilde H}, \hat{\mathbf z}'],
\end{equation}
with $\hat{\tilde H} \equiv \hat{\mathbf z}'^\dag \tilde{\mathbf H}
\hat{\mathbf z}'$,

\begin{eqnarray}
&&  \hat{\tilde H} =   -\Delta \hat \Xi^\dag \hat \Xi 
  + W \bigl(\hat\Phi^\dag\hat\Xi + \hat\Xi^\dag\hat\Phi\bigr) 
  - i \dot\theta \bigl(\hat\Psi^\dag \hat\Phi - \hat\Phi^\dag \hat\Psi\bigr)
  % \\+ \frac {\hat\sigma_{11} + \hat \sigma_{33} -1}N
  % [-\Delta\hat\sigma_{33} + (\Omega\hat\sigma_{31} + \mbox{H.c.})
  ,\nonumber \\
&&  \tilde{\mathbf H} = {\mathbf R} {\mathbf H} {\mathbf R}^{-1} 
  +i {\dot {\mathbf R}}{\mathbf R}^{-1}
  =
  \begin{pmatrix}
    0 & -i\dot\theta & 0 \\
    i\dot\theta  & 0 & W \\
    0 & W & -\Delta
  \end{pmatrix},
  \label{eq:Heit2}
\end{eqnarray} 
where $W \equiv \sqrt{g^2N_A+\Omega^2}$.

We immediately recognize that in the adiabatic limit $\dot\theta \ll
W$, the dark-state polaritons are decoupled from the dynamics.
Moreover, they do not involve the excited atomic state $\ket3$ and
hence are immune to spontaneous emission.  Therefore, if the initial
state of the system consists of dark-state polaritons only and the
mixing angle $\theta$ varies slowly enough, then the quantum state of
the system adiabatically follows the smoothly changing dark
eigenstates of the Hamiltonian, so the system stays in the same
superposition of dark states as it was initially in.  (Note that the
zero eigenvalue is nondegenerate in each invariant subspace $\mathcal
S_n$, so there is no level crossing.)  Thus, adiabatic rotation of the
mixing angle $\theta$ from $0$ to $\pi/2$ leads to a complete transfer
of the photonic state to collective atomic excitations between levels
$\ket1$ and $\ket2$.  Indeed, for $\theta=0$, the dark-state polariton
mode $\hat\Psi(\theta)$ coincides with the signal mode $\hat a_L$,
while it solely corresponds to the collective spin excitation $\hat
\sigma_{21} /\sqrt{N_A}$ for $\theta=\pi/2$.  This is the standard way
of interpreting the adiabatic EIT storage process~\citep{prl84p5094,
  pra65e022314}.

The bright-state polaritons are coupled to the excited-state mode that
decays by spontaneous emission with rate $\Gamma$.  If this relaxation
process is fast enough ($W\ll\Gamma$) or the one-photon resonance is
far detuned ($W\ll\Delta$), we can eliminate the excited mode.  Then
we find an effective decay $\gamma_B$ of the bright-state polaritons
and a shift $\omega_B$ in their energy,
\begin{equation}
  \label{eq:eit-omegaB}
  \omega_B = \frac{W^2 \Delta}{\Gamma^2/4+\Delta^2},
  \qquad
  \gamma_B = \frac{W^2 \Gamma}{\Gamma^2/4+\Delta^2}.
\end{equation}

%%%%%%%%%%%%%%%%%%%%%%%%%%%%%%%%%%%%%%%%%%%%%%%%%%%%%%%
\subsubsection{Far off-resonant regime}

Now we identify two important limiting cases.  In the first, the
one-photon resonance is far detuned with respect to the Rabi frequency
$\Omega$ of the control field, to the ensemble-enhanced vacuum Rabi
frequency $g\sqrt{N_A}$ of the signal field, and also to the decay
rate $\Gamma$ of the excited states.  In this case ($W\ll\Delta$ and
$\Gamma\ll\Delta$), the bright energy shift $\omega_B$ dominates over
the decay,
\begin{equation}
  \gamma_B = \omega_B \frac\Gamma\Delta
  \ll \omega_B = \frac{W^2}\Delta.
  \label{eq:omegaB-off}
\end{equation}
Switching back to Heisenberg picture with non-rotating light and
atomic variables, we obtain the following Hamilton operator,
\begin{multline}
  \label{eq:Heit-off}
  \hat H = \frac{g^2N_A}\Delta \hat a_L^\dag \hat a_L
  + \frac{\Omega^2}\Delta \frac{\hat\sigma_{12} \hat\sigma_{21}}{N_A}
  \\
  + \frac{g\sqrt{N_A}\Omega}\Delta \left(
    \hat a_L^\dag \frac{\hat\sigma_{21}}{\sqrt{N_A}}
    + \frac{\hat\sigma_{12}}{\sqrt{N_A}} \hat a_L^\dag \right).
\end{multline}
The first two terms are AC Stark shifts for the signal mode
interacting with the atoms in level $\ket2$ and for atoms in level
$\ket1$ interacting with the intense control field, respectively.
These terms were eliminated in \sref{sec:Raman} by considering a
manifold of excited states and tuning the laser fields to a point
where their AC Stark contributions exactly compensate.  Finally, the
third term in \eqref{eq:Heit-off} coincides with \eqref{eq:HLambda4}
if we take $\Omega=g|\alpha|$ (the factor~2 is due to the fact that
there we had two upper levels).  Finally we mention that the
characteristic time $T$ of the write process is determined by
\eqref{eq:intpi2} and is of the order $T \sim \Delta /W^2$.
\Eref{eq:omegaB-off} then implies that accumulated losses are of the
order $\gamma_B T \sim \Gamma/\Delta$ and can be safely neglected.

%%%%%%%%%%%%%%%%%%%%%%%%%%%%%%%%%%%%%%%%%%%%%%%%%%%%%%%
\subsubsection{Resonant regime}

The decay of bright-state polaritons cannot be neglected near
one-photon resonance.  Instead, we obtain an effective Liouvillian for
the rotating density operator $\hat\rho'$ of the dark and
bright polariton modes,
\begin{equation}
  \label{eq:eit-master}
  \frac{d}{dt}\hat\rho' 
  = -i \bigl[\hat{\tilde H}_{\text{eff}},\hat\rho'\bigr]
  + {\mathcal L}_{\text{diss}} \hat\rho'.
\end{equation}
Here the effective Hamiltonian in the rotating axes representation is
given by
\begin{equation}
  \label{eq:eit-Heff}
  \hat{\tilde H}_{\text{eff}} = \omega_B {\hat \Phi}^\dagger {\hat \Phi}
  -i {\dot \theta} \bigl(\hat\Psi^\dag \hat\Phi
  -\hat\Phi^\dagger \hat\Psi\bigr),
\end{equation}
and the dissipative Lindblad superoperator is
\begin{equation}
  \label{eq:eit-lindblad}
  {\mathcal L}_{\text{diss}} \hat\rho = 
    \frac{\gamma_B}2
    (2\hat\Phi \hat\rho \hat\Phi^\dag
    - \hat\Phi^\dag \hat\Phi \hat\rho
    - \hat\rho \hat\Phi^\dag \hat\Phi).
\end{equation}

If the adiabatic condition $\dot\theta \ll\gamma_B$ is satisfied,
terms with $\dot\theta$ can be neglected, so the dark and bright
polaritons become adiabatically decoupled.  This means that in the
rotating axes representation, both the dark and bright polaritons can
be considered as free quasi-particles (except that the latter have
finite lifetime).  If the initial state contains no bright polaritons
at all---that is, the atomic ensemble is totally polarized at
$\theta=0$ in the beginning of the write process (or the signal light
mode is in the vacuum state at $\theta= \frac\pi2$ in the beginning of
the read-out process)---the Lindblad term does not play a role. As a
consequence, the dynamics is entirely described by the Hamiltonian
\eqref{eq:eit-Heff} that can be considered zero for exact one-photon
resonance ($\Delta=0$).  However, when we change from the rotating
axes representation back to Heisenberg picture, the term neglected in
the adiabatic approximation reappears.  In the original basis of light
and matter operators, this translates to
\begin{gather}
  \nonumber
  \mathbf H_{\text{eff}} = {\mathbf R}^{-1} 
  \tilde{\mathbf H}_{\text{eff}} {\mathbf R} 
  -i {\dot {\mathbf R}}{\mathbf R}^{-1}
  =
  \begin{pmatrix}
    0 &  i\dot\theta \\
    -i\dot\theta & 0
  \end{pmatrix},
\end{gather}
and thus to 
\begin{gather}
  \hat{H}_{\text{eff}} = i \dot\theta
  (\hat a_L^\dag \hat\sigma_{21} 
  - \hat a_L \hat\sigma_{12} )/{\textstyle\sqrt{N_A}}
  .\label{eq:eit-Heff2}
\end{gather} 
Introducing the atomic and light quadrature variables as
\begin{align}
  \hat X_A &\equiv (\hat\sigma_{21} + \hat\sigma_{12}) 
  /\sqrt{2N_A},&
  \hat X_L &\equiv (\hat a_L + \hat a_L^\dag) /\sqrt2,
  \nonumber\\
  \hat P_A &\equiv (\hat\sigma_{21} - \hat\sigma_{12}) 
  /i\sqrt{2N_A},&
  \hat P_L &\equiv (\hat a_L - \hat a_L^\dag) /i\sqrt2,
  \label{eq:eit-quadratures}
\end{align}
we find that the effective Hamiltonian is 
\begin{equation}
  \hat H_{\text{eff}} =
  %% \frac12 \omega_B \cos^2\theta (\hat X_A^2 + \hat P_A^2)
  %% + \frac12 \omega_B \sin^2\theta (\hat X_L^2 + \hat P_L^2)
  %% \\ + \omega_B \sin\theta\cos\theta (\hat X_A \hat X_L + \hat P_A \hat P_L)
  %% +
  \dot\theta (\hat X_A \hat P_L - \hat P_A \hat X_L),
  \label{eq:eit-HadXP}
\end{equation}
and we recover the ideal mapping Hamiltonian \eqref{eq:Hideal1}. 
Complete mapping from light to atoms or vice versa 
is achieved if
the time dependence of the mixing angle is adjusted such that
\begin{equation}
  \pm\frac\pi2 = \int_0^T d\tau\, \dot\theta (\tau)
  = \theta(T) - \theta(0).
\end{equation}

Finally we mention that non-adiabatic losses of dark-state polaritons
are characterized by the decay rate
\begin{equation}
  \label{eq:eit-gammaD}
  \gamma_D = \frac{\dot\theta^2}{\gamma_B} \ll \dot\theta,
\end{equation}
where we applied the adiabatic condition $\dot\theta \ll\gamma_B$.
As the characteristic time $T$ of the write\slash read-out process is
of the order $T \sim 1/\dot\theta$, the accumulated losses are of the
order $T\gamma_D \sim \gamma_B / \dot\theta \ll 1$.

%%%%%%%%%%%%%%%%%%%%%%%%%%%%%%%%%%%%%%%%%%%%%%%%%%%%%%%%%%%%%%%%%%
%                                                                %
%                           Section                              %
%                                                                %
%%%%%%%%%%%%%%%%%%%%%%%%%%%%%%%%%%%%%%%%%%%%%%%%%%%%%%%%%%%%%%%%%%
\section{Conclusions}
\label{sec:conclusions}

In the present paper we have investigated quantum memory for light in
atomic ensemble using a continuous variable description, i.e.,
position and momentum variables for light and matter degrees of
freedom.  In particular we have studied in detail two types of
off-resonant quantum memories: one based on the quantum Faraday effect
supplemented by measurement and feedback, and another involving
$\Lambda$-type atoms and Raman scattering.  For the first scheme we
analyzed the effect of inefficiencies in the initial atomic state
preparation and imperfections in the light measurement. We found that
the fidelity of the one-pass memory scheme for storing superpositions
of light states with very different momenta is rather low unless
atomic state preparation can be performed with very high accuracy. In
a two-pass scheme the necessity of an initial atomic spin squeezing
can be avoided by light measurement and feedback. In a triple-pass
configuration, atomic state preparation as well as feedback can be
avoided altogether.  Secondly, we have investigated an off-resonant
Raman scheme and proposed a configuration in which unwanted light
shifts can be cancelled. Finally we have discussed near resonant
quantum memories based on electromagnetically induced transparency in
terms of position and momentum operators of light and matter. We have
shown that this memory can be described by a Hamiltonian corresponding
to an ideal map provided that before the write and read-out processes
the atomic ensemble or, respectively, the radiation mode are in the
appropriate initial vacuum state.  In contrast to the common approach
involving wave equations for the propagating polaritons, we have
developed a Hamiltonian formalism for the atomic and photonic
quadrature variables similar to the formalism of the first two
families.  Our results allow a straightforward comparison of the
various continuous variable quantum memory schemes in the same
framework.

\begin{acknowledgments}
  The financial support from the EMALI network of the European
  Commission (Contract No.  MRTN-CT-2006-035369) is gratefully
  acknowledged.
\end{acknowledgments}

\bibliographystyle{myfullaps}
\bibliography{qmem-refs}
\end{document}